\documentclass[12pt]{article}
\usepackage{multicol}
\usepackage{amssymb}
\usepackage{amsfonts,amssymb,amsmath,amsthm}
\usepackage{epsfig}
\usepackage{slashed}
\usepackage{graphicx}
\usepackage{subfig}
\usepackage{mathrsfs}
\usepackage{bbm}
\usepackage{cite}
\usepackage{enumerate}
\usepackage{slashbox}
\usepackage{color}
\usepackage[colorlinks,linkcolor=red,anchorcolor=red,citecolor=green]{hyperref}
\makeatletter 
\@addtoreset{equation}{section}
\makeatother  

\makeatletter

\newcommand{\Rmnum}[1]{\expandafter\@slowromancap\romannumeral #1@}
\makeatother

\begin{document}
\renewcommand{\thefootnote}{\fnsymbol{footnote}}
\begin{titlepage}

\vspace{10mm}
\begin{center}
{\Large\bf Validity of Maxwell Equal Area Law for Black Holes Conformally Coupled to Scalar Fields in $\text{AdS}_5$ Spacetime}
\vspace{16mm}

{{\large Yan-Gang Miao${}^{}$\footnote{\em E-mail: miaoyg@nankai.edu.cn}
and Zhen-Ming Xu}${}^{}$\footnote{\em E-mail: xuzhenm@mail.nankai.edu.cn}

\vspace{6mm}
${}^{}${\normalsize \em School of Physics, Nankai University, Tianjin 300071, China}

}

\end{center}

\vspace{10mm}
\centerline{{\bf{Abstract}}}
\vspace{6mm}
We investigate the $P-V$ criticality and the Maxwell equal area law for a five-dimensional spherically symmetric AdS black hole with a scalar hair in the absence of and in the presence of a Maxwell field, respectively. Especially in the charged case, we give the exact $P-V$ critical values. More importantly, we analyze the validity and invalidity of the Maxwell equal area law for the AdS hairy black hole in the scenarios without and with charges, respectively. Within the scope of validity of the Maxwell equal area law, we point out that there exists a representative van der Waals-type oscillation in the $P-V$ diagram. This oscillating part that indicates the phase transition from a small black hole to a large one can be replaced by an isobar. The small and large black holes share the same Gibbs free energy. We also give the distribution of the critical points in the parameter space both without and with charges, and obtain for the uncharged case the fitting formula of the co-existence curve. Meanwhile, the latent heat is calculated, which gives the energy released or absorbed between the small and large black hole phases in the isothermal-isobaric procedure.
\vskip 20pt
\noindent
{\bf PACS Number(s)}: 04.50.Gh, 04.70.-s, 04.70.Dy


\end{titlepage}

\newpage
\renewcommand{\thefootnote}{\arabic{footnote}}
\setcounter{footnote}{0}
\setcounter{page}{2}
\pagenumbering{arabic}
\tableofcontents
\vspace{1cm}

\section{Introduction}
Since the seminal works by Hawking and Bekenstein on the radiation of black holes, the exploration of thermodynamic properties of black holes has received a wide range of attention~\cite{JMB,RW,SC} and also acquired great progress~\cite{CGKP,AKMS,BC,MSSM}. Of more particular interest is the thermodynamics of anti-de Sitter (AdS) black holes~\cite{SH,RGC} where the AdS/CFT duality plays a pivotal role in recent developments of theoretical physics~\cite{RD,BPD3}. In the context of AdS/CFT correspondence~\cite{JMM,EW}, the Hawking-Page phase transition~\cite{WP} of five-dimensional AdS black holes can be explained as the phenomenon of the confinement/deconfinement transition in the four-dimensional Yang-Mills gauge field theory~\cite{EW1}. Another archetypical example of the AdS/CFT correspondence is the holographic superconductor, which can be regarded as the scalar field condensation around a four-dimensional charged AdS black hole~\cite{HHH}.

With the cosmological constant $\Lambda$ being treated as a thermodynamic pressure variable~\cite{BPD,CEJM,KM,DSJT,BPD2,HPPFM} and its conjugate variable being considered as the thermodynamic volume, the thermodynamics in the extended phase space has been getting more and more attentions.  In this paradigm, the mass of black holes is identified as the enthalpy rather than the internal energy. This idea has also been applied to other known parameters, such as the Born-Infeld parameter~\cite{GKM,NB}, the Gauss-Bonnet coupling constant~\cite{RCLY}, the noncommutative parameter~\cite{YGMX}, and the Horndeski non-minimal kinetic coupling strength~\cite{YGMX1}, etc. All the parameters just mentioned can be regarded as a kind of thermodynamic pressure. Furthermore, there exists a similar situation in the exploration of charged AdS hairy black holes~\cite{HM} of Einstein-Maxwell theory conformally coupled to a scalar field in five dimensions. The model's action has been given by~\cite{GLOR,GGO,GGGO,MC}
\begin{equation}
\emph{I}=\frac{1}{\kappa}\int \text{d}^5 x\sqrt{-g}\left(R-2\Lambda-\frac14 F^2+\kappa \textit{L}_{m}(\phi, \nabla\phi)\right), \label{action}
\end{equation}
where $\kappa=16\pi G$,  $G$ is the five-dimensional Newton constant, $R$ the scalar curvature, $F$ the electromagnetic field strength, $g_{\mu\nu}$ the metric with mostly plus signatures, and $g=\text{det}(g_{\mu\nu})$. In addition, the Lagrangian matter $\textit{L}_{m}(\phi, \nabla\phi)$ takes the form in five dimensions as follows,
\begin{eqnarray}
\textit{L}_{m}(\phi, \nabla\phi)= b_0 \phi^{15}+b_1 \phi^7 {S_{\mu\nu}}^{\mu\nu}+b_2 \phi^{-1}\left({S_{\mu\lambda}}^{\mu\lambda}{S_{\nu\delta}}^{\nu\delta}
-4{S_{\mu\lambda}}^{\nu\lambda}{S_{\nu\delta}}^{\mu\delta}+{S_{\mu\nu}}^{\lambda\delta}{S^{\nu\mu}}_{\lambda\delta}\right), \label{lag}
\end{eqnarray}
where $b_0$, $b_1$, and $b_2$ are coupling constants and the four-rank tensor
\begin{eqnarray}
{S_{\mu\nu}}^{\lambda\delta}=\phi^2{R_{\mu\nu}}^{\lambda\delta}-12\delta_{[\mu}^{[\lambda}\delta_{\nu]}^{\delta]}\nabla_{\rho}\phi\nabla^{\rho}\phi
 -48\phi \delta_{[\mu}^{[\lambda}\nabla_{\nu]}\nabla^{\delta]}\phi+18\delta_{[\mu}^{[\lambda}\nabla_{\nu]}\phi \nabla^{\delta]}\phi,
\end{eqnarray}
has been shown~\cite{GGO,GGGO,MC,OR} to transform covariantly, ${S_{\mu\nu}}^{\lambda\delta}\rightarrow \Omega^{-8/3}{S_{\mu\nu}}^{\lambda\delta}$, under the Weyl transformation, $g_{\mu\nu}\rightarrow \Omega^2 g_{\mu\nu}$ and $\phi\rightarrow \Omega^{-1/3}\phi$.

In fact, one can see that the above model is the most general scalar field/gravity coupling formulation whose field equations are of second order for both gravity and matter. Hence, we can say that this formulation is a generalization of the Horndeski theory~\cite{GWH} whose action contains a non-minimal kinetic coupling of a massless real scalar field and the Einstein tensor. Even more importantly, being a simple and tractable model, it provides a significant advantage for studying the phase transition of hairy black holes on the AdS spacetime where the back-reaction of the scalar field on the metric can be solved analytically in five dimensions. In the paradigm where there exists a complete physical analogy between the four-dimensional Reissner-Nordstr\"{o}m AdS black hole and the real van der Waals fluid~\cite{KM} in the phase transition, the recent research~\cite{HM} shows that this charged AdS hairy black hole also exhibits the van der Waals-type thermodynamic behavior, and moreover, such a black hole undergoes reentrant phase transition which usually occurs in higher curvature gravity theory. We note that all the interesting results just mentioned are available by making the coupling parameter dynamical, i.e. treating the coupling parameter as a certain thermodynamic variable.

Based on the results pointed out above, we take advantage of the well-established Maxwell equal area law~\cite{ESAS,WL,PHN,CNP,DMS,XX} to make a further investigation of the van der Waals-type phase transition, of the co-existence curve,  and of the $P-V$ critical phenomenon for this charged AdS hairy black hole. We give analytically the critical values of the charged hairy black hole and show in detail the behavior of the phase transition from a small black hole to a large one. We also give the distribution of the critical points in the parameter space of $q$ (the coupling constant of the scalar field) and $e$ (the electric charge of the black hole), and obtain for the uncharged case the fitting formula of the co-existence curve. More importantly, we analyze the validity and invalidity of the Maxwell equal area law for the five-dimensional charged AdS hairy black hole and determine the conditions of the law holding. Meanwhile, the latent heat is calculated, which gives the energy released or absorbed between the small and large black hole phases in the isothermal-isobaric procedure.

The paper is organized as follows. In section \ref{sec2}, we review the analytic solution of the charged AdS hairy black hole in $D=5$ dimensions and some relevant thermodynamic quantities. In section \ref{sec3}, we calculate the $P-V$ critical values and investigate the Maxwell equal area law for this five-dimensional charged AdS hairy black hole. This section contains two subsections which correspond to the scenarios without and with charges, respectively. Finally, we devote to drawing our conclusion in section \ref{sec4}.

\section{Analytic solution in $D=5$ dimensions}\label{sec2}
The model described by eqs.~(\ref{action}) and (\ref{lag}) admits~\cite{HM,GGGO} an exactly electrically charged solution in five dimensions, 
\begin{equation}
\text{d}s^2=-f \text{d}t^2+\frac{\text{d}r^2}{f}+r^2 \text{d}\Omega^2_{3(k)},
\end{equation}
where the function $f$ takes the form,
\begin{equation}
f(r)=k-\frac{m}{r^2}-\frac{q}{r^3}+\frac{e^2}{r^4}+\frac{r^2}{l^2},
\end{equation}
$\text{d}\Omega^2_{3(k)}$ is the metric of the three-dimensional surface with a constant curvature, the curvature is positive for $k=1$, zero for $k=0$, and negative for $k=-1$, and $m$ and $e$ are two integration constants corresponding to the mass and the electric charge of the black hole, respectively. Here the parameter $l$ represents the curvature radius of the AdS spacetime, which is associated with the cosmological constant $\Lambda$ whose role is analogous to the thermodynamic pressure,
\begin{equation}\label{pressure}
P=-\frac{\Lambda}{8\pi}=\frac{3}{4\pi l^2}.
\end{equation}
Moreover, the parameter $q$ is characterized as the coupling constant of the scalar field, 
\begin{equation}
q=\frac{64\pi}{5}\varepsilon k b_1\left(-\frac{18kb_1}{5b_0}\right)^{3/2}, \label{defq}
\end{equation}
where $\varepsilon=-1,0,1$ and there exists an additional constraint: $10b_0 b_2=9b_1^2$, to ensure the existence of this black hole solution. These conditions imply that $q$ only takes values $0, \pm |q|$. Meanwhile, the scalar field configuration takes the form,
\begin{equation}
\phi(r)=\frac{n}{r^{1/3}}, \qquad n=\varepsilon\left(-\frac{18kb_1}{5b_0}\right)^{1/6},
\end{equation}
and the Maxwell gauge potential reads
\begin{equation}
A=\frac{\sqrt{3}e}{r^2},
\end{equation}
where the field strength still takes the standard form: $F_{\mu\nu}=\partial_{\mu}A_{\nu}-\partial_{\nu}A_{\mu}$.

The location of this hairy black hole horizon is denoted by $r_h$, which is taken to be the largest real positive root of $f(r)=0$. That is, the horizon radius $r_h$ satisfies the following polynomial equation~\cite{GGGO},
\begin{equation}
r_h^6+k l^2 r_h^4-ml^2 r_h^2-ql^2 r_h+e^2 l^2=0.
\end{equation}

For the horizon thermodynamic properties of this hairy black hole, some thermodynamic quantities have been calculated in refs.~\cite{HM,GGGO} and are listed below for our later use. The thermodynamic enthalpy $M$, temperature $T$, entropy $S$, and charge $Q$ take the following forms,
\begin{eqnarray}
M&=&\frac{3\omega_{3(k)}}{16\pi}m=\frac{3\omega_{3(k)}}{16\pi}\left(kr_h^2-\frac{q}{r_h}+\frac{e^2}{r_h^2}+\frac{r_h^4}{l^2}\right), \label{enth}\\
T&=&\frac{k}{2\pi r_h}+\frac{q}{4\pi r_h^4}-\frac{e^2}{2\pi r_h^5}+\frac{r_h}{\pi l^2}, \label{temp}\\
S&=&\frac{\omega_{3(k)}}{4}\left(r_h^3-\frac{5}{2}q\right), \label{entr}\\
Q&=&-\frac{\omega_{3(k)}\sqrt{3}}{16\pi}e, \label{char}
\end{eqnarray}
where $\omega_{3(k)}$ denotes the area of the compact three-dimensional manifold with the metric ${\text d}\Omega^2_{3(k)}$. In order to develop the first law and the Smarr relation, the coupling parameter is dealt with~\cite{HM} as a dynamical variable, which means that $q$ is extended to be a continuous and real parameter. Thus, $q$ should appear in the Smarr relation and its variation should be included in the first law of thermodynamics to make the first law of black hole thermodynamics be consistent with the Smarr relation.

Overall, the extended first law of thermodynamics can be written in terms of the thermodynamic quantities mentioned above as follows,
\begin{equation}
\text{d}M=T \text{d}S+V\text{d}P+\Phi \text{d}Q+K \text{d}q, \label{law}
\end{equation}
where the thermodynamic volume $V$,  the electric potential $\Phi$, and the extensive variable $K$ conjugate to the coupling parameter $q$ have the forms,
\begin{eqnarray}
V&\equiv&\left(\frac{\partial M}{\partial P}\right)_{S,Q,q}=\frac{\omega_{3(k)}}{4}r_h^4, \label{volu}\\
\Phi&\equiv&\left(\frac{\partial M}{\partial Q}\right)_{S,P,q}=-\frac{2\sqrt{3}}{r_h^2}e, \label{elepo}\\
K&\equiv&\left(\frac{\partial M}{\partial q}\right)_{S,P,Q}=\frac{\omega_{3(k)}}{32 l^2 r_h^5}\left[20r_h^6+2r_h^4 l^2\left(5k-3\right)+5qr_h l^2-10e^2 l^2\right]. \label{extenq}
\end{eqnarray}
Correspondingly, the extended Smarr relation can be deduced~\cite{HM},
\begin{equation}
2M=3T S-2PV+2\Phi Q+3qK. \label{smarr}
\end{equation}
Meanwhile, the Gibbs free energy and the equation of state for this hairy black hole can be written in terms of eqs.~(\ref{enth}), (\ref{temp}), and (\ref{entr}) as follows,
\begin{eqnarray}
G&\equiv&M-TS \nonumber\\
&=&\frac{\omega_{3(k)}}{16\pi}\left\{kr_h^2-\frac{r_h^4}{l^2}+\frac{5q^2}{2r_h^4}
\left(\frac{10r_h}{l^2}+\frac{5k-4}{r_h}\right)q+\left(\frac{5}{r_h^2}-\frac{5q}{r_h^5}\right)e^2\right\}, \label{gfree}\\
P(r_h, T)&=&\frac{3T}{4r_h}-\frac{3k}{8\pi r_h^2}-\frac{3q}{16\pi r_h^5}+\frac{3e^2}{8\pi r_h^6}. \label{state}
\end{eqnarray}

Next, we shall discuss the Maxwell equal area law for this hairy black hole in order to investigate its critical behavior and the co-existence curve of two phases resorting to eqs.~(\ref{volu}) and (\ref{state}) in the $(P, V)$ plane. Note that in the planar case, i.e. $k=0$, according to eq.~(\ref{defq}), we can see $q=0$. It implies that there are no hairs. In addition, for the hyperbolic case, i.e. $k=-1$, there are no physically critical values as pointed out by ref.~\cite{HM}.
Hence, we shall focus only on the spherical case, i.e. $k=1$, in the following context.

\section{Maxwell equal area law}\label{sec3}
The thermodynamic behavior of black holes in the AdS background is analogous to that of the real van der Waals fluid. As was  known, the critical behavior of the van der Waals fluid occurs at the critical isotherm $T=T_c$ when the $P-V$ diagram has an inflection point,
\begin{equation}
\frac{\partial P}{\partial r_h}=0, \qquad \frac{\partial^2 P}{\partial r_h^2}=0. \label{creq}
\end{equation}
When $T<T_c$, there is an oscillating part in the $P-V$ diagram. We have to replace this oscillating part by an isobar in order to describe it in such a way that the areas above and below the isobar are equal to each other. This treatment is based on the Maxwell equal area law. Thus, by making an analogy between the black hole in the AdS background and the real van der Waals fluid, we find that there also exists an oscillating part below the critical temperature $T_c$ in the $P-V$ diagram, which indicates that the first order phase transition occurs from a small black hole to a large one. This isobar that satisfies the Maxwell equal area law represents the co-existence curve of small and large black holes~\cite{XX}.

Normally, the Maxwell equal area law is constructed in the $(P, V)$ plane for a constant temperature, and it can also be made in the $(T, S)$ or $(\Phi, Q)$ plane. These constructions are equivalent. Theoretically, the law can be established from the variation of the Gibbs free energy defined by eq.~(\ref{gfree}),
\begin{equation}
\text{d} G=\text{d} M-T \text{d}S-S \text{d}T. \label{dgibbs}
\end{equation}
With resorting to the first law of thermodynamics~(\ref{law}) and keeping in mind that the co-existing phases have the same Gibbs free energy, one thus arrives at the Maxwell equal area law in the $(P, V)$ plane by integrating eq.~(\ref{dgibbs}) at constant $T$, $Q$, and $q$,
\begin{equation}
\begin{aligned}
&P(r_1, T)=P(r_2, T)=P^{*},\\
&P^{*}\cdot(V_2-V_1)=\int_{r_1}^{r_2}P(r_h, T)\,\text{d}V, \label{arealaw}
\end{aligned}
\end{equation}
where $P^{*}$ stands for an isobar, $V_1$ and $V_2$ denote the thermodynamic volume defined by eq.~(\ref{volu}) for the small and large black holes with the horizon radii $r_1$ and $r_2$, respectively. Thanks to the Maxwell equal area law eq.~(\ref{arealaw}), we can also obtain the latent heat which represents the amount of energy released or absorbed from one phase to the other in the isothermal-isobaric condition,
\begin{equation}
L=T\left[S_2-S_1\right], \label{latent}
\end{equation}
where $S_1$ and $S_2$ denote the entropy defined by eq.~(\ref{entr}) for the small and large black holes with the horizon radii $r_1$ and $r_2$, respectively. It is worth mentioning that the co-existence curve of the two phases, i.e. the small and large black holes that are described by an isobar in the Maxwell equal area law, is governed by the Clausius-Clapeyron equation,
\begin{equation}
\left(\frac{\text{d}P}{\text{d}T}\right)_{Q,q}=\frac{S_2-S_1}{V_2-V_1}. \label{cceq}
\end{equation}
In the following we specialize in the Maxwell equal area law, the critical values, and the co-existing phases for this AdS hairy black hole.

\subsection{Uncharged case: $e=0$}
For this hairy black hole with the spherical symmetry in the absence of a Maxwell field, according to eqs.~(\ref{state}) and (\ref{gfree}), we can see that the equation of state reduces to be
\begin{equation}
P(r_h, T)=\frac{3T}{4r_h}-\frac{3}{8\pi r_h^2}-\frac{3q}{16\pi r_h^5}, \label{state1}
\end{equation}
and that the Gibbs free energy becomes\footnote{In the spherical case, i.e. $k=1$, the area of a compact three-dimensional manifold $\omega_{3(k)}$ equals $2 \pi^2$.}
\begin{eqnarray}
G=\frac{\pi}{8}\left\{r_h^2-\frac{r_h^4}{l^2}+\frac{5q^2}{2r_h^4}+\left(\frac{10r_h}{l^2}+\frac{1}{r_h}\right)q\right\}. \label{gfree1}
\end{eqnarray}

\subsubsection{Critical values}
The critical values defined by eq.~(\ref{creq}) have been obtained in ref.~\cite{HM},
\begin{eqnarray}
r_c&=& (-5q)^{1/3}, \nonumber \\
T_c&=&-\frac{3}{20}\cdot \frac{(-5q)^{2/3}}{\pi q}, \nonumber\\
P_c&=&\frac{9}{200\pi}\left(-\frac{\sqrt{5}}{q}\right)^{2/3}, \label{c1}
\end{eqnarray}
which exist only for $q<0$, and the relevant ratio reads
\begin{equation}
\frac{P_c r_c}{T_c}=\frac{3}{10}. \label{r1}
\end{equation}
We observe that this ratio is a constant that does not depend on the parameter $q$, and that it is different from that of the real van der Waals fluid or the five-dimensional Reissner-Nordstr\"{o}m AdS black hole~\cite{KM}.

\subsubsection{Maxwell equal area law}

Inserting eqs.~(\ref{state1}) and (\ref{volu}) into eq.~(\ref{arealaw}),
we give\footnote{For the details of derivation, see Appendix A.} the Maxwell equal area law,
\begin{equation}
\begin{aligned}
&2r_1^3 r_2^3 \left(r_1+r_2\right)+q\left[\left(r_1+r_2\right)^4+4r_1^2 r_2^2\right]=0, \\
&4\pi T r_1^3 r_2^3\left(r_1^2+r_1 r_2+r_2^2\right)+3q\left[\left(r_1+r_2\right)^4-r_1^2 r_2^2\right]=0. \label{arealaw1}
\end{aligned}
\end{equation}
When taking the critical limit $r_1=r_2=r_c$, we can see that eq.~(\ref{arealaw1}) turns back to eq.~(\ref{c1}). Moreover, with the help of eqs.~(\ref{entr}) and (\ref{arealaw1}) we obtain the latent heat $L$ between the small and large black hole phases,
\begin{equation}
L=\frac{\pi^2 T}{2}\left(r_2^3-r_1^3\right). \label{latent1}
\end{equation}

\begin{figure}
\begin{center}
  \begin{tabular}{ccc}
    \includegraphics[width=50mm]{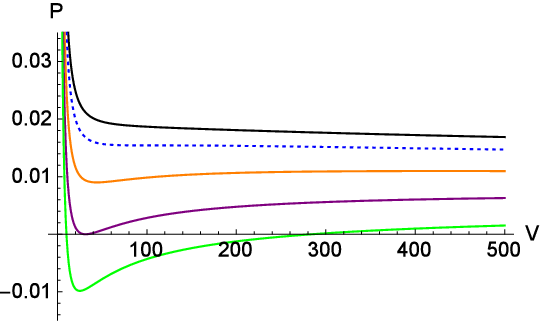} &
    \includegraphics[width=50mm]{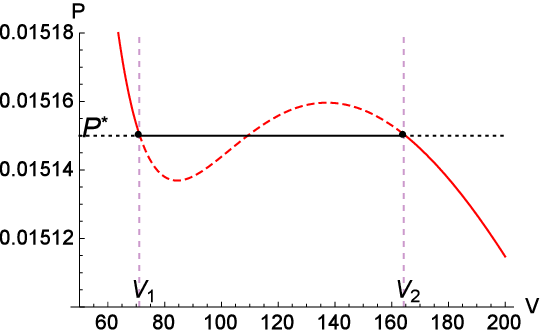} &
    \includegraphics[width=50mm]{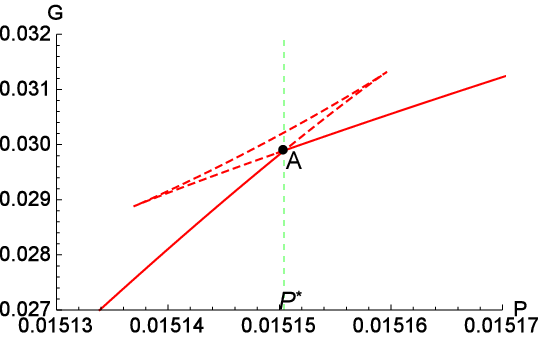} \\
  \end{tabular}
\end{center}
\caption{(color online) $P-V$ diagrams described by eqs.~(\ref{state1}) and (\ref{volu}) and $P-G$ diagram described by eq.~(\ref{gfree1}) at $q=-2$. \textbf{Left}: The temperature takes the values of $T=0.12000$ $(\text{black})$, $T=T_c=0.11081$ $(\text{blue dotted})$, $T=0.09500$ $(\text{orange})$, $T=0.07520$ $(\text{purple})$, and $T=0.05500$ $(\text{green})$, respectively. \textbf{Middle}: The temperature takes the value of $T=0.11000$, and for the isobar $P^{*}=0.01515$, $V_1$ and $V_2$ correspond to $r_1=1.94812$ and $r_2=2.40186$, respectively. Corresponding to the middle diagram, the Gibbs free energy is depicted in \textbf{Right} diagram, showing the characteristic swallowtail behaviour.}
\label{pv1}
\end{figure}

In order to highlight the outstanding thermodynamic properties of the uncharged case, the numerical calculations about the Maxwell equal area law eq.~(\ref{arealaw1}) in the $(P, V)$ plane are displayed in Table \ref{biao1}, meanwhile the $P-V$ critical behavior and the Gibbs free energy described by eqs.~(\ref{state1}),~(\ref{gfree1}), and (\ref{volu}) are portrayed in Figure \ref{pv1} for the specific value of $q=-2$. From Table \ref{biao1}, we can see clearly that the horizon radii of the small and large black holes, $r_1$ and $r_2$, shrink into the critical horizon radius $r_c=2.15443$ when the isotherm condition $T=T_c=0.11081$ is taken. It implies that no phase transitions occur and correspondingly the latent heat $L$ is of course equal to zero. At this moment, the isobar $P^{*}$ becomes the critical pressure $P_c=0.01543$. With gradually decreasing of the temperature $T$ from the critical value $T_c=0.11081$, it is clear that $r_1$ decreases while $r_2$ increases, and the Maxwell equal area law is always valid. Furthermore, the small black hole with the radius $r_1$ and the large one with the radius $r_2$ share the same Gibbs free energy, where the values of the Gibbs free energy are listed in the fifth column of Table \ref{biao1}. Quantitatively, the latent heat $L$ between the two phases takes a sharp increase with the temperature decreasing. Qualitatively, our analysis is given below. As we have known, $r_1$ decreases while $r_2$ increases when $T$ decreases. Combining eq.~(\ref{arealaw1}) with eq.~(\ref{latent1}), we can get $L \sim \frac{3\pi}{4}r_2^2$ when $T$ is decreasing. Due to $r_2$ increasing, the latent heat $L$ increases in the quadratic form of $r_2$. As a result, the qualitative analysis  coincides with the quantitative one shown in Table \ref{biao1}.

\begin{table}[!hbp]
\begin{center}
\begin{tabular}{|c|*{6}{|c}}
\hline
\multicolumn{6}{|c|} {$q=-2$} \\ \hline
$T$       & $r_1$     & $r_2$     & $P^{*}$   & $G$ & $L$ \\ \hline \hline
0.11081   & $2.15443$ & $2.15443$ & $0.01543$ & $0$ & $0$ \\
0.11041   & $2.00562$ & $2.32350$ & $0.01529$ & $0.01478$ & $2.43883$ \\
0.11000   & $1.94812$ & $2.40186$ & $0.01515$ & $0.02989$ & $3.50814$ \\
0.10500   & $1.67534$ & $2.93675$ & $0.01352$ & $0.21142$ & $10.6873$ \\
0.09500   & $1.46590$ & $3.72579$ & $0.01069$ & $0.56174$ & $22.7697$ \\
0.08000   & $1.30792$ & $4.96216$ & $0.00728$ & $1.06476$ & $47.3528$ \\
0.07520   & $1.27272$ & $5.41946$ & $0.00637$ & $1.22154$ & $58.3034$ \\
0.05500   & $1.16325$ & $8.02830$ & $0.00329$ & $1.86510$ & $140.017$\\ \hline
\end{tabular}
\end{center}
\caption{The numerical results of the Maxwell equal area law eqs.~(\ref{arealaw}) and (\ref{arealaw1}), the Gibbs free energy eq.~(\ref{gfree1}), and the latent heat eq.~(\ref{latent1}) for the hairy black hole with the spherical symmetry in the absence of a Maxwell field.}
\label{biao1}
\end{table}

Next, we take a close look at Figure \ref{pv1} which portrays the critical behavior of the uncharged case. The left diagram demonstrates the representative $P-V$ critical curves. When the temperature $T$ exceeds the critical temperature $T_c$, see the black curve, there is no any criticality. The middle diagram illustrates the typical equal area law, where we indeed observe a van der Waals-type oscillation shown in the red curve. By analogy with the real van der Waals fluid, this oscillating part must be replaced by an isobar in the $P-V$ diagram. Adopting the Maxwell equal area law, i.e., the areas of the regions surrounded by an oscillation (red dashed curve) and an isobar (black solid line) are equal to each other, we can find out  this isobar $P^{*}$ and correspondingly the thermodynamic volumes $V_1$ and $V_2$. In other words, the phase transition from the small black hole to the large one occurs in this situation, and these two types of black holes share the same Gibbs free energy displayed by the crossing point $A$ in the right diagram.

\begin{figure}
\begin{center}
  \begin{tabular}{cc}
    \includegraphics[width=60mm]{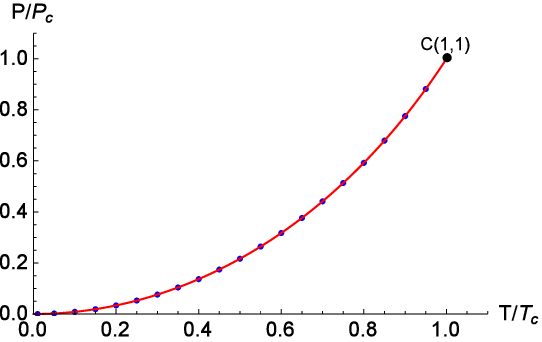} &
    \includegraphics[width=60mm]{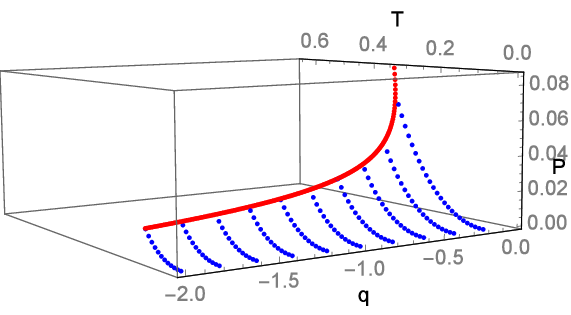} \\
  \end{tabular}
\end{center}
\caption{(color online) \textbf{Left}: The co-existence curve for the five-dimensional AdS hairy black hole without charges, the blue points  for the numerical values, and the red solid curve for the fitting formula. \textbf{Right}: The co-existence curve in the parameter $q$ space. Each blue dotted curve indicates a co-existence curve at a fixed $q$ and the red curve at the boundary corresponds to the critical points.}
\label{pt1}
\end{figure}

With the aid of the Clausius-Clapeyron equation eq.~(\ref{cceq}) and the Maxwell equal area law eq.~(\ref{arealaw}), we can obtain the co-existence curve of two phases. Considering that the co-existence curve for the real van der Waals fluid has a positive slope everywhere and terminates at the critical point, we try to fit it using a polynomial. Thanks to the simple forms of the critical point eq.~(\ref{c1}), such a treatment can be realized. By  introducing the reduced parameters,
\begin{equation}
t\equiv \frac{T}{T_c}, \qquad p\equiv \frac{P}{P_c}, \label{rps}
\end{equation}
we eventually obtain the parametrization form of the co-existence curve,
\begin{eqnarray}\label{fitcoline}
p &=& 0.828355 t^2+0.096202 t^3-0.558203 t^4+2.43626 t^5-4.74242 t^6 \nonumber \\
  & & +5.05509 t^7-2.53495 t^8+0.241581 t^9+0.178083 t^{10},
\end{eqnarray}
where $t\in (0,1)$. Then we plot the numerical values governed by the Clausius-Clapeyron equation eq.~(\ref{cceq}) and the Maxwell equal area law eq.~(\ref{arealaw}), and also plot the fitting formula eq.~(\ref{fitcoline}) of the co-existence curve of the small and large black holes in Figure \ref{pt1}. We can see clearly that the numerical values and the fitting formula match well with each other. Both of them terminate at the critical point, i.e. the point $C(1,1)$. With the increasing of $q$, the critical values, see eq.~(\ref{c1}), take an increasing trend shown in the right diagram of Figure \ref{pt1}.

At last, we make a brief summary of this subsection. For the AdS hairy black hole with the spherical symmetry in the absence of a Maxwell field, the $P-V$ critical behavior is available only under the condition of $q<0$, and the Maxwell equal area law always holds under this condition. Meanwhile, we take the specific value of $q=-2$ as an example to highlight the salient features of this black hole in Table \ref{biao1} and Figure \ref{pv1}. Furthermore, because the critical point eq.~(\ref{c1}) has a simple form, we obtain the fitting formula~(\ref{fitcoline}) of the co-existence curve, see Figure \ref{pt1} for the illustration. As to the charged case, it has a little complexity that is exhibited in the following subsection.

\subsection{Charged case: $e\neq 0$}
For the spherically symmetric AdS hairy black hole with the Maxwell field, the equation of state takes the form,
\begin{equation}
P(r_h, T)=\frac{3T}{4r_h}-\frac{3}{8\pi r_h^2}-\frac{3q}{16\pi r_h^5}+\frac{3e^2}{8\pi r_h^6}, \label{state2}
\end{equation}
and the Gibbs free energy reads
\begin{eqnarray}
G=\frac{\pi}{8}\left\{r_h^2-\frac{r_h^4}{l^2}+\frac{5q^2}{2r_h^4}+\left(\frac{10r_h}{l^2}+\frac{1}{r_h}\right)q
+\left(\frac{5}{r_h^2}-\frac{5q}{r_h^5}\right)e^2\right\}. \label{gfree2}
\end{eqnarray}

\subsubsection{Critical values}
Substituting eq.~(\ref{state2}) into eq.~(\ref{creq}), we can obtain the critical radius which satisfies the following equation,
\begin{equation}
r_c^4+5q r_c-15e^2=0, \label{crrad}
\end{equation}
and derive the critical temperature,
\begin{equation}
T_c=\frac{3(16e^2-5q r_c)}{4\pi r_c^5}, \label{crtem}
\end{equation}
and the critical pressure,
\begin{equation}
P_c=\frac{3(10e^2-3q r_c)}{8\pi r_c^6}. \label{crpre}
\end{equation}
Then, using eqs.~(\ref{crrad}), (\ref{crtem}), and (\ref{crpre}), we give the ratio of three critical values,
\begin{equation}
\frac{P_c r_c}{T_c}=\frac{10e^2-3q r_c}{2(16e^2-5qr_c)}, \label{r2}
\end{equation}
which turns back to eq.~(\ref{r1}), i.e. the uncharged case, when $e=0$. Moreover, if $q=0$, this ratio equals $5/16$, which coincides with the ratio of the five-dimensional Reissner-Nordstr\"{o}m AdS black hole~\cite{KM}. Nonetheless, this ratio depends in general on the values of parameters $e$ and $q$.

Let us try to solve eq.~(\ref{crrad}). At first, we set $w(r_c)\equiv r_c^4+5q r_c-15e^2$ and take a look at its asymptotic behavior. When $r_c \rightarrow 0$, we get a negative value, $w(r_c)\rightarrow -15e^2$. When $r_c \rightarrow +\infty$, we see $w(r_c)\rightarrow +\infty$. In addition, we notice that for $q>0$ the function $w(r_c)$ is monotone increasing at the interval $[0, +\infty)$, and for $q<0$ it is monotone increasing at the interval $[(-5q/4)^{1/3}, +\infty)$ but monotone decreasing at the interval $[0,(-5q/4)^{1/3}]$. Hence, we can conclude that the equation $w(r_c)=0$ must have one and only one positive root,\footnote{For the details of derivation, see Appendix B.} 
\begin{equation}
r_c=\frac{1}{2}\left\{\left(\frac{10|q|}{\sqrt{t_c}}-t_c\right)^{1/2}\pm \sqrt{t_c}\right\}, \label{root}
\end{equation}
where the plus sign corresponds to the case of $q<0$ and the minus sign to the case of $q>0$, and the newly introduced parameter $t_c$ is defined as
\begin{equation*}
t_c \equiv 4\sqrt{5}e \sinh\left\{\frac13 \sinh^{-1}\left(\frac{\sqrt{5} q^2}{16e^3}\right)\right\}.
\end{equation*}

\subsubsection{Maxwell equal area law}
Substituting eqs.~(\ref{state2}) and (\ref{volu}) into eq.~(\ref{arealaw}), we give\footnote{For the details of calculations, see Appendix A.} the Maxwell equal area law,
\begin{equation}
\begin{aligned}
&2r_1^4 r_2^4 \left(r_1+r_2\right)+qr_1 r_2\left[\left(r_1+r_2\right)^4+4r_1^2 r_2^2\right]-2e^2\left[\left(r_1+r_2\right)^5-r_1 r_2\left(r_1^3+r_2^3\right)\right]=0, \\
&4\pi T r_1^4 r_2^4\left(r_1^2+r_1 r_2+r_2^2\right)+3qr_1 r_2\left[\left(r_1+r_2\right)^4-r_1^2 r_2^2\right]-6e^2\left(r_1+r_2\right)^3\left(r_1^2+r_1 r_2+r_2^2\right)=0. \label{arealaw2}
\end{aligned}
\end{equation}
If $e=0$,  i.e. for the uncharged case, the above set of equations turns back to eq.~(\ref{arealaw1}).
When taking the critical limit, $r_1=r_2=r_c$, we can see that eq.~(\ref{arealaw2}) reduces to the eqs.~(\ref{crrad}) and (\ref{crtem}), that is, it coincides with the critical behavior. It is remarkable for the charged case that the entropy defined by eq.~(\ref{entr}) probably turns into negative in the case of $q>0$, which brings an extra constraint on the $P-V$ criticality and the equal area law.

\subsubsection*{\Rmnum{1}. The case of $q<0$}

In this situation, the entropy of the hairy black hole with the Maxwell field, eq.~(\ref{entr}), is always positive. There are no additional restrictive conditions for investigating the Maxwell equal area law and the phase transition. The treatment is the same as that of the scenario without charges. We can directly write down the condition under which the van der Waals-type phase transition exists and the Maxwell equal area law holds, i.e., the temperature takes the values $T <T_c$, where $T_c$ is given by eq.~(\ref{crtem}) together with  the constraint $q<0$.

\begin{figure}
\begin{center}
  \begin{tabular}{ccc}
    \includegraphics[width=50mm]{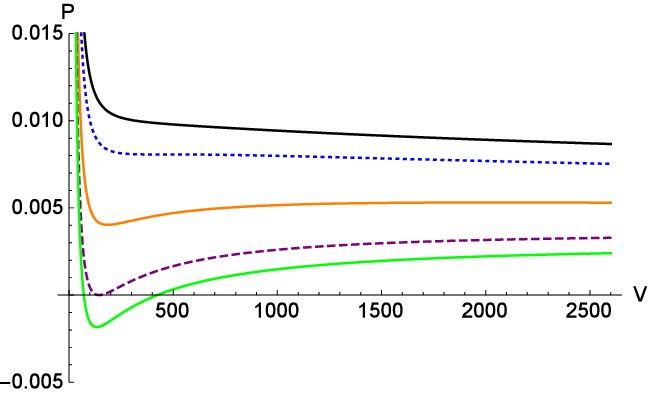} &
    \includegraphics[width=50mm]{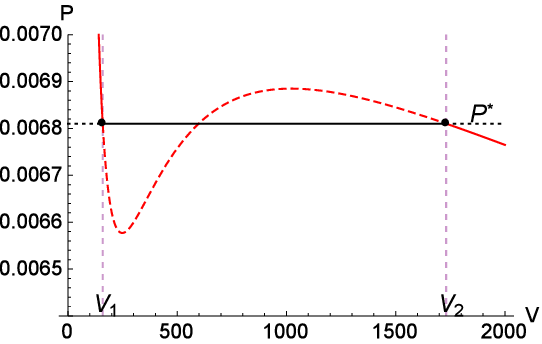} &
    \includegraphics[width=50mm]{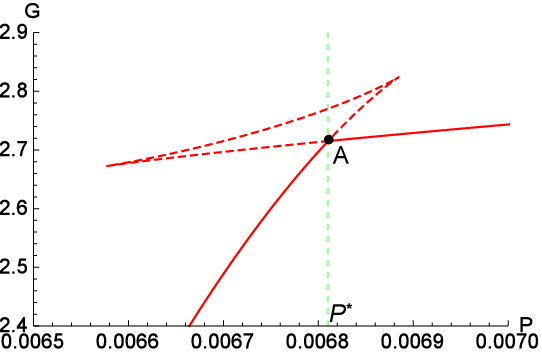} \\
  \end{tabular}
\end{center}
\caption{(color online) $P-V$ diagrams described by eqs.~(\ref{state2}) and (\ref{volu}) at $q=-2$ and $e=2$. \textbf{Left}: The temperature takes the values of $T=0.08800$ $(\text{black})$, $T=T_c=0.08072$ $(\text{blue dotted})$, $T=0.06650$ $(\text{orange})$,  $T=0.05364$ $(\text{purple dashed})$, and $T=0.04800$ $(\text{green})$, respectively. \textbf{Middle}: The temperature takes the value of $T=0.07520$,  and for the isobar $P^{*}=0.00681$, $V_1$ and $V_2$ correspond to $r_1=2.38204$ and $r_2=4.32768$, respectively. Corresponding to the middle diagram, the Gibbs free energy described by eq.~(\ref{gfree2}) is depicted in \textbf{Right} diagram, showing the characteristic swallowtail behaviour.}
\label{pv2}
\end{figure}

Table \ref{biao2} displays the numerical results of the Maxwell equal area law (eq.~(\ref{arealaw2})), and Figure \ref{pv2} depicts the critical behavior of the equation of state (eqs.~(\ref{state2}) and (\ref{volu})) and the Gibbs free energy (eq.~(\ref{gfree2})) for an example of the charged hairy black hole  at $q=-2$ and $e=2$. We can see the similar critical behavior to that of the uncharged case. When $T>T_c$, there is no criticality; when $T <T_c$, there exists a representative van der Waals-type oscillation in the $P-V$ diagram. By using the Maxwell equal area law and replacing the oscillating part by an isobar, we observe the phase transition from the small black hole to the large one in the middle diagram of Figure \ref{pv2}, and find that the two phases share the same Gibbs free energy displayed by the crossing point A in the right diagram. Moreover, when $T$ is decreasing, the horizon radius of the small black hole, $r_1$, decreases while that of the large black hole, $r_2$, increases, and the latent heat $L$ between the two phases presents a sharp increasing tendency. This observation means that the phase transition needs more energy at the lower temperature of the isothermal-isobaric procedure.

\begin{table}[!hbp]
\begin{center}
\begin{tabular}{|c|*{6}{|c}}
\hline
\multicolumn{6}{|c|} {$q=-2$ and $e=2$} \\ \hline
$T$       & $r_1$     & $r_2$     & $P^{*}$   & $G$ & $L$ \\ \hline \hline
0.08072   & $3.08753$ & $3.08753$ & $0.00806$ & $2.38767$ & $0$ \\
0.07520   & $2.38204$ & $4.32768$ & $0.00681$ & $2.71511$ & $25.0625$ \\
0.07050   & $2.20615$ & $5.00649$ & $0.00587$ & $2.98087$ & $39.9219$ \\
0.06650   & $2.10367$ & $5.58115$ & $0.00514$ & $3.19951$ & $53.9958$ \\
0.06020   & $1.98519$ & $6.54903$ & $0.00413$ & $3.53222$ & $81.1202$ \\
0.05640   & $1.92919$ & $7.19679$ & $0.00358$ & $3.72703$ & $101.746$ \\
0.05364   & $1.89349$ & $7.70977$ & $0.00322$ & $3.86627$ & $119.502$ \\
0.04800   & $1.83056$ & $8.90086$ & $0.00254$ & $4.14463$ & $165.582$ \\ \hline
\end{tabular}
\end{center}
\caption{The numerical results of the Maxwell equal area law eqs.~(\ref{arealaw}) and (\ref{arealaw2}), the Gibbs free energy eq.~(\ref{gfree2}), and the latent heat eq.~(\ref{latent}) for the spherically symmetric AdS hairy black hole with the Maxwell field.}
\label{biao2}
\end{table}

\subsubsection*{\Rmnum{2}. The case of $q>0$}
In the situation of a positive parameter $q$, the analysis becomes complicated. Due to the entropy described by eq.~(\ref{entr}) being probably negative, we have to impose an extra constraint to the horizon radius,
\begin{equation}
r_h \geq r_s \equiv \left(\frac{5q}{2}\right)^{1/3}. \label{excon}
\end{equation}
The purpose is to avoid the negative entropy that appears if $r_h < r_s$,  where the negative entropy is regarded as an unphysical variable at present.

The Maxwell equal area law holds under the condition $T <T_c$ together with the constraint eq.~(\ref{excon}), where $T_c$ is given by eq.~(\ref{crtem}) under the condition $q>0$.

\begin{figure}
\begin{center}
  \begin{tabular}{cc}
    \includegraphics[width=60mm]{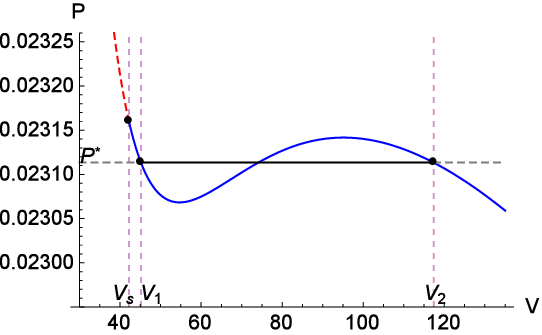} &
    \includegraphics[width=60mm]{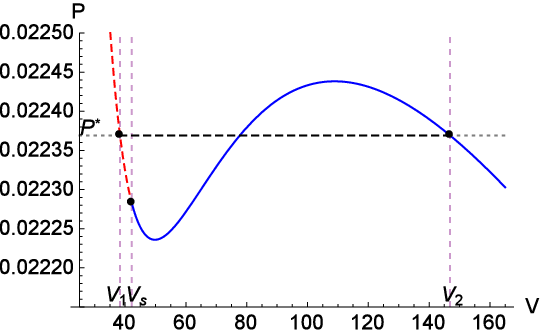} \\
    \includegraphics[width=60mm]{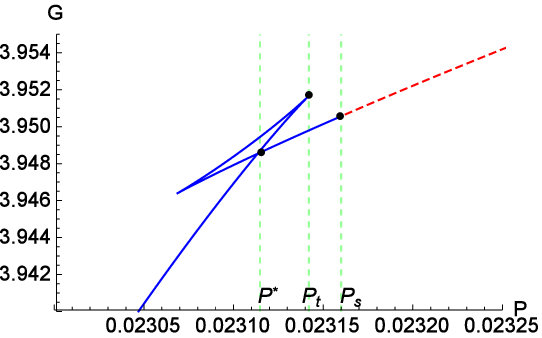} &
    \includegraphics[width=60mm]{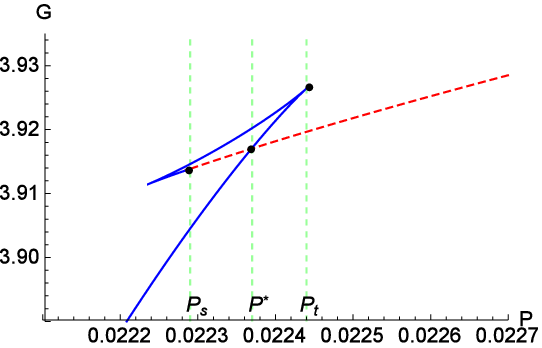} \\
  \end{tabular}
\end{center}
\caption{(color online) $P-V$ diagrams (eqs.~(\ref{state2}) and (\ref{volu})) and the corresponding Gibbs free energy (eq.~(\ref{gfree2})) at $q=2$ and $e=1.5$. \textbf{Left}: The temperature takes the value of $T=0.14000$. For the isobar $P^{*}=0.02311$, $V_s$, $V_1$, and $V_2$ correspond to $r_s=1.70998$, $r_1=1.73921$, and $r_2=2.20835$, respectively. Note that $r_1 > r_s$. \textbf{Right}: The temperature takes the value of $T=0.13800$. For the isobar $P^{*}=0.02237$, $V_s$, $V_1$, and $V_2$ correspond to $r_s=1.70998$, $r_1=1.66931$, and $r_2=2.33574$, respectively. Note that $r_1 < r_s$.}
\label{pv3}
\end{figure}

Now we consider an extreme situation, i.e., $r_c=r_s$, which leads to the critical value of the electric charge,
\begin{equation}
|e_s|=\frac{\sqrt{5}}{5}\left(\frac{5q}{2}\right)^{2/3}=\frac{r_s^2}{\sqrt{5}}. \label{ecrit}
\end{equation}
This critical value ($r_s$ or $e_s$) gives the boundary of violating or maintaining the $P-V$ criticality and the Maxwell equal area law.
That is, if $|e|<|e_s|$, we have $r_c<r_s$, resulting in the violation of the $P-V$ criticality (eqs.~(\ref{crrad})-(\ref{crpre})) and of the Maxwell equal area law (eq.~(\ref{arealaw2})); if $|e|>|e_s|$, we have $r_c>r_s$, maintaining the validity of the $P-V$ critical values, but  we have to add the condition $r_1>r_s$ in order to establish the Maxwell equal area law. A sample of the latter situation is depicted in Figure \ref{pv3} for the specific fixing of $q=2$ and $e=1.5$. In this sample, we have $e>e_s=1.30766$ and figure out $r_s=1.70998$ and $r_c=1.94471$. Hence, the validity of the Maxwell equal area law depends on the temperature with the range of $T <T_c$, where $T_c=0.14207$. The left two diagrams of Figure \ref{pv3} correspond to  $T=0.14000 < T_c=0.14207$ that leads to $r_1>r_s$, and illustrates the equal area law holding and the characteristic swallowtail behavior existing. The red dashed curve corresponds to the negative entropy which is unphysical, but it does not affect the application for the equal area law and the characteristic swallowtail structure owing to $r_1>r_s$. On the contrary, if $r_1<r_s$, see the right two diagrams of Figure \ref{pv3}, the red dashed curve  corresponding to the negative entropy indeed violates the equal area law and the characteristic swallowtail structure. Let us take a close look at this kind of  violation. We can still determine the isobar $P^{*}$ because the initial point at $r_1$ (or $V_1$) and the terminal one at $r_2$ (or $V_2$) are independent of eq.~(\ref{excon}).  However, when $T=0.13800$ that brings about $r_1<r_s$, i.e. the breaking of eq.~(\ref{excon}), we find that the branch of negative entropy terminates at $r_s$ (or $V_s$) that is larger than $r_1$, which evidently leads to the violation of the Maxwell equal area law and the characteristic swallowtail structure.


\begin{figure}
\begin{center}
  \begin{tabular}{ccc}
    \includegraphics[width=50mm]{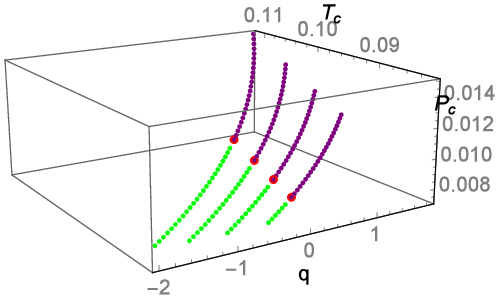} &
    \includegraphics[width=53mm]{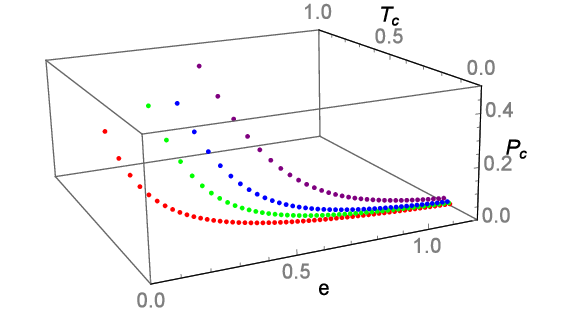} &
    \includegraphics[width=52mm]{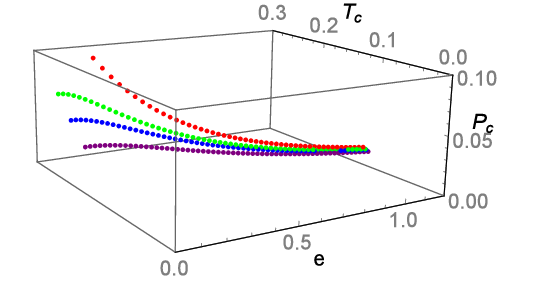} \\
  \end{tabular}
\end{center}
\caption{(color online) The distribution of the critical points described by eqs.~(\ref{crtem}) and (\ref{crpre}) for different values of $e$ and $q$. \textbf{Left} diagram corresponds to $e=1.90$, $2.05$, $2.20$, and $2.35$, respectively, from top to bottom, the green dotted curve for $q<0$, the purple dotted curve for $q>0$, and the red dots for $q=0$. \textbf{Middle} diagram depicts the case of $q \ge 0$ with discrete values of $q=0$ (\text{red}), $0.3$ (\text{green}), $0.6$ (\text{blue}), and $1.2$ (\text{purple}), and \textbf{Right} diagram depicts the case of $q \le 0$ with discrete values of $q=0$ (\text{red}), $-0.2$ (\text{green}), $-0.3$ (\text{blue}), and $-0.5$ (\text{purple}).}
\label{pt2}
\end{figure}

In addition, we take a look at the co-existence curve governed by the Clausius-Clapeyron equation eq.~(\ref{cceq}) and the Maxwell equal area law eq.~(\ref{arealaw2}) for the five-dimensional AdS hairy black hole with charges. As the formulas of the critical values, eqs.~(\ref{crtem}) and (\ref{crpre}), are much more complicated than eq.~(\ref{c1}), it is hard to give a perfect fitting formula of the co-existence curve of the small and large black holes. Nonetheless, we can know through the above analysis that the thermodynamic behavior for the case with charges is the same as that of the case without charges within the scope of validity of the Maxwell equal area law, and that the co-existence curve always has a positive slope and terminates at the critical point. 
Figure \ref{pt2} shows the critical points for different values of $e$ and $q$, from which we can see that the critical values $T_c$ and $P_c$ take an increasing trend with an increasing $q$ but a fixed $e$, while they take a decreasing trend with an increasing $e$ but a  fixed $q$.

\begin{figure}
\begin{center}
    \includegraphics[width=80mm]{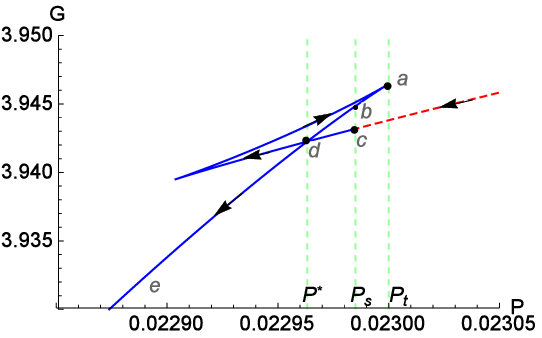}
\end{center}
\caption{(color online) The Gibbs free energy (eq.~(\ref{gfree2})) at $q=2$, $e=1.5$, and $T=0.13960$. $P^{*}$ corresponds to the pressure of the intersection (point d), $P_{t}$ corresponds to the pressure of the cusp (point a), and $P_{s}$ corresponds to the pressure of point c with zero entropy. The red dashed curve corresponds to the negative entropy. A black arrow indicates an increasing trend of the horizon radius $r_h$.}
\label{pv4}
\end{figure}

Before the end of this section, let us take a look at the physical meaning of the appearance of negative entropy in the charged case with positive $q$. We illustrate this situation by setting $q=2$ and $e=1.5$, where the behavior of the Gibbs free energy (eq.~(\ref{gfree2})) can be classified into three types that are shown in Figure \ref{pv4} and the left-down and right-down diagrams of Figure \ref{pv3}. We know that the stable black hole is thermodynamically favorable to the lower Gibbs free energy.
\begin{itemize}
  \item If $P^{*} <P_s<P_t$, see Figure \ref{pv4}, one can see that the global minimum of the Gibbs free energy emerges a discontinuous characteristic due to the negative entropy (see the red dashed curve). More precisely, when $P_s <P<P_t$, the global minimum is  located on the curve $a\rightarrow b$. When $P^{*} <P<P_s$, the global minimum is located on the curve $c\rightarrow d$. When $P<P^{*}$, the global minimum is located on the curve $d\rightarrow e$. At $P=P_s$, the Gibbs free energy has a discontinuous global minimum, which means the occurrence of the zeroth order phase transition as reported in ref.~\cite{HM}. At $P=P^{*}$, the global minimum of the Gibbs free energy has an inflection point, which implies a standard first order small/large black hole phase transition, i.e. the van der Waals-type phase transition, and the Maxwell equal area law holds.
  \item If $P_s > P_t$, see the left-down diagram of Figure \ref{pv3}, the global minimum of the Gibbs free energy is continuous and also has an inflection point (corresponding to the pressure $P^{*}$), indicating the existence of the van der Waals-type phase transition.
  \item  If $P_s< P^{*}$, see the right-down diagram of Figure \ref{pv3}, the global minimum of the Gibbs free energy is continuous but has no inflection points, which indicates nonexistence of  phase transitions and results in invalidity of the Maxwell equal area law.
  \end{itemize}
All in all, it is the positivity/negativity of the entropy that affects  the behavior of the global minimum of the Gibbs free energy. The discontinuity of the Gibbs free energy causes a zeroth order phase transition at the pressure $P_s\in [P^{*},P_t]$ under a certain temperature. Moreover, if $P_s<P^{*}$, the disappearance of inflection points leads to the failure of the Maxwell equal area law, so that no real physical isobar exists for phase transitions, as shown in the right-top diagram of Figure \ref{pv3}.

\section{Conclusion}\label{sec4}
In this paper, we investigate the $P-V$ criticality, the Maxwell equal area law, and the co-existence curve for the spherically symmetric AdS black hole with a scalar hair~\cite{HM,GGGO} both in the absence of and in the presence of a Maxwell field, respectively. Especially in the charged case, we give the exactly analytical $P-V$ critical values, see eqs.~(\ref{crrad}), (\ref{crtem}), (\ref{crpre}), and (\ref{root}). Meanwhile, we provide the conditions of validity of the Maxwell equal area law for the hairy black hole without and with charges, respectively. Our results can be summarized as follows:
\begin{itemize}
  \item The case of $q<0$
  \begin{itemize}
    \item Scenario without charges: the Maxwell equal area law holds under the conditions $T<T_c$ and eq.~(\ref{c1}).
    \item Scenario with charges: the Maxwell equal area law holds under the conditions $T<T_c$ and eq.~(\ref{crtem}).
  \end{itemize}
  \item The case of $q>0$
  \begin{itemize}
    \item Scenario without charges: the Maxwell equal area law is violated.
    \item Scenario with charges: besides the conditions $T<T_c$ and eq.~(\ref{crtem}) together with the extra constraint eq.~(\ref{excon}), whether the Maxwell equal area law holds or not depends on the charge and the relation between the temperature and the horizon radius, which can be classified into the following two situations,
        \begin{itemize}
          \item when $|e|<|e_s|$, the law does not hold.
          \item when $|e|>|e_s|$, the law holds if $r_1 > r_s$, but fails if $r_1 < r_s$, where $r_1$ depends on the temperature $T$ shown in Figure \ref{pv3}.
        \end{itemize}
  \end{itemize}
\end{itemize}

Within the scope in which the Maxwell equal area law holds, we point out that there exists a representative van der Waals-type oscillation in the $P-V$ diagram. This oscillating part that indicates the phase transition from a small black hole to a large one can be replaced by an isobar and the small and large black holes share the same Gibbs free energy. These salient features have been illustrated in Tables \ref{biao1} and \ref{biao2}, and Figures \ref{pv1} and \ref{pv2}, from which we conclude that when the temperature $T$ is decreasing, the horizon radius of the small black hole $r_1$ decreases while that of the large black hole $r_2$ increases, and moreover, the latent heat $L$ between the two phases presents a sharp increasing tendency. This observation means that the phase transition needs more energy at the lower temperature of the isothermal-isobaric procedure. Furthermore, for the uncharged case we obtain the fitting formula~(\ref{fitcoline}) of the co-existence curve depicted in Figure \ref{pt1} due to the simple form of the critical point eq.~(\ref{c1}). For the charged case, we give the distribution of the critical points described by eqs.~(\ref{crtem}) and (\ref{crpre}) in the parameter space of $q$ and $e$ in Figure \ref{pt2}. Finally, we point out that the positivity/negativity of the entropy makes effect on the global minimum of the Gibbs free energy. The inflection point of the Gibbs free energy leads to the van der Waals-type phase transition, but the Maxwell equal area law does not hold and the real physical isobar does not exist for phase transitions if no inflection points exist as shown in Figure \ref{pv3}. The discontinuity of the Gibbs free energy causes a zeroth order phase transition as shown in Figure \ref{pv4}.

\section*{Acknowledgments}
This work was supported in part by the National Natural Science Foundation of China under grant No.11675081. At last, the authors would like to thank the anonymous referee for the helpful comment that indeed greatly improves this work.

\section*{Appendix}
\subsection*{A. Derivation of Eq.~(\ref{arealaw1}) and Eq.~(\ref{arealaw2})}

Inserting eqs.~(\ref{state}) and (\ref{volu}) into eq.~(\ref{arealaw}), we can obtain the following three equations,
\begin{align}
&\frac{3T}{4r_1}-\frac{3k}{8\pi r_1^2}-\frac{3q}{16\pi r_1^5}+\frac{3e^2}{8\pi r_1^6}=P^{*}, \tag{A1} \label{a1}\\
&\frac{3T}{4r_2}-\frac{3k}{8\pi r_2^2}-\frac{3q}{16\pi r_2^5}+\frac{3e^2}{8\pi r_2^6}=P^{*}, \tag{A2} \label{a2}\\
&\frac{\omega_{3(k)}P^{*}}{4}(r_2^4-r_1^4)=\int_{r_1}^{r_2}\left[\frac{3T}{4r_h}-\frac{3k}{8\pi r_h^2}-\frac{3q}{16\pi r_h^5}+\frac{3e^2}{8\pi r_h^6}\right]\text{d}\left(\frac{\omega_{3(k)}}{4}r_h^4\right).\tag{A3} \label{a3}
\end{align}
Combining eq.~(\ref{a1}) with eq.~(\ref{a2}), we have
\begin{equation}
4\pi Tr_1^5 r_2^5\left(r_2-r_1\right)+2e^2\left(r_2^6-r_1^6\right)-qr_1r_2\left(r_2^5-r_1^5\right)-2kr_1^4r_2^4\left(r_2^2-r_1^2\right)=0, \tag{A4} \label{a4}
\end{equation}
and inserting eqs.~(\ref{a1}) and (\ref{a2}) into eq.~(\ref{a3}) yields
\begin{equation}
4\pi Tr_1^2 r_2^2\left(r_2^3-r_1^3\right)+18e^2\left(r_2^2-r_1^2\right)-15qr_1r_2\left(r_2-r_1\right)-6kr_1^2r_2^2\left(r_2^2-r_1^2\right)=0. \tag{A5} \label{a5}
\end{equation}
Comparing eq.~(\ref{a4}) with eq.~(\ref{a5}) and eliminating the parameter $T$, we obtain
\begin{equation}
2kr_1^4 r_2^4 \left(r_1+r_2\right)+qr_1 r_2\left[\left(r_1+r_2\right)^4+4r_1^2 r_2^2\right]-2e^2\left[\left(r_1+r_2\right)^5-r_1 r_2\left(r_1^3+r_2^3\right)\right]=0. \tag{A6} \label{a6}
\end{equation}
Substituting eq.~(\ref{a6}) into eq.~(\ref{a5}) and eliminating the term containing the parameter $k$, we get
\begin{equation}
4\pi T r_1^4 r_2^4\left(r_1^2+r_1 r_2+r_2^2\right)+3qr_1 r_2\left[\left(r_1+r_2\right)^4-r_1^2 r_2^2\right]-6e^2\left(r_1+r_2\right)^3\left(r_1^2+r_1 r_2+r_2^2\right)=0. \tag{A7} \label{a7}
\end{equation}
For the case of $k=1$, eqs.~(\ref{a6}) and (\ref{a7}) turn out to be eq.~(\ref{arealaw2}), and eq.~(\ref{arealaw2}) reduces to eq.~(\ref{arealaw1}) when $e=0$ is taken.

\subsection*{B. Root of Eq.~(\ref{crrad})}

For a special quartic equation, such as $x^4+bx-c^2=0$, we set
\begin{equation}
x^4+bx-c^2=(x^2+\gamma x+\alpha)(x^2-\gamma x+\beta), \tag{B1} \label{b1}
\end{equation}
where $\alpha$, $\beta$, and $\gamma$ are parameters to be determined. At first, we write the four roots of the above quartic equation by solving the two quadratic equations separated from the quartic one,
\begin{equation}
x_{1,2}=\frac{-\gamma\pm \sqrt{\gamma^2-4\alpha}}{2}, \qquad x_{3,4}=\frac{\gamma\pm \sqrt{\gamma^2-4\beta}}{2}. \tag{B2} \label{b2}
\end{equation}
Next, expanding the right hand side of eq.~(\ref{b1}) and comparing the terms with the same power to $x$, we have
\begin{align}
\alpha+\beta &= \gamma^2, \tag{B3} \label{b3}\\
-\alpha+\beta &=\frac{b}{\gamma}, \tag{B4} \label{b4}\\
\alpha \beta &=-c^2. \tag{B5} \label{b5}
\end{align}
Solving eqs.~(\ref{b3}) and (\ref{b4}), we first determine two of the three parameters, $\alpha$ and $\beta$,
\begin{equation}
\alpha=\frac12 \left({\gamma}^2-\frac{b}{\gamma}\right), \qquad \beta=\frac12 \left({\gamma}^2+\frac{b}{\gamma}\right). \tag{B6} \label{b6}
\end{equation}
In order to fix the parameter $\gamma$, we then insert eq.~(\ref{b6}) into eq.~(\ref{b5}) and introduce the new parameter $y \equiv \gamma^2$, which  can change the sixth-order equation with respect to $\gamma$ into a cubic equation with respect to the new parameter $y$,
\begin{equation}
y^3+4c^2 y-b^2=0. \tag{B7} \label{b7}
\end{equation}
For this kind of cubic equations, there is only one real root that is given by a hyperbolic form of Vi\`{e}te's solution,
\begin{equation}
y=\frac{4\sqrt{3}}{3}c \sinh\left\{\frac13 \sinh^{-1}\left(\frac{3\sqrt{3} b^2}{16c^3}\right)\right\}, \tag{B8} \label{b8}
\end{equation}
which actually determines the last parameter $\gamma$.
As a result, eqs.~(\ref{b2}), (\ref{b6}), and (\ref{b8}) give the exactly analytical roots of the original quartic equation.


\end{document}